\documentclass[conf]{new-aiaa}
\usepackage[utf8]{inputenc}

\usepackage{graphicx}
\usepackage{amsmath}
\usepackage[usenames,dvipsnames]{xcolor}
\usepackage{colortbl}
\usepackage[version=4]{mhchem}
\usepackage{siunitx}
\usepackage{longtable,tabularx}
\title{Pressure Reconstruction from the Measured Pressure Gradient Using Gaussian Process Regression}

\author{Zejian You\footnote{Doctoral student, Department of Aerospace Engineering, San Diego State University, Student Member AIAA.}  and Qi Wang\footnote{Assistant Professor, Department of Aerospace Engineering, San Diego State University, Member AIAA. E-mail: qwang4@sdsu.edu}}
\affil{San Diego State University, San Diego, California, 92182, USA}
\author{Xiaofeng Liu\footnote{Associate Professor, Department of Aerospace Engineering, San Diego State University, Associate Fellow AIAA.}}
\affil{San Diego State University, San Diego, California, 92182, USA}

\begin{document}

\maketitle

\begin{abstract}
Many numerical algorithms have been established to reconstruct pressure fields from measured kinematic data with noise by Particle Image Velocimetry (PIV), such as the Pressure Poisson solver and the Omni-Directional Integration (ODI) method. This study adopts Gaussian Process Regression (GPR), a probabilistic framework with an intrinsic de-noising mechanism to tackle drawbacks of traditional Pressure Poisson solver and compares the performance with ODI. To evaluate the accuracy of the algorithm, GPR and ODI are tested in detail in a canonical setup of forced homogeneous isotropic turbulence from the Johns Hopkins Turbulence Database. According to the result, GPR has the same level of accuracy as ODI with optimized hyper-parameters for the isotropic turbulence flow. However, GPR has the tendency to flatten impulsive signals. Therefore, without further modifications, it is not suitable to detect flow structures with impulsive true signals. The error propagation of the proposed framework is also analyzed and discussed in both physical and spectral spaces.
\end{abstract}

\section*{Nomenclature}

{\renewcommand\arraystretch{1.0}
\noindent\begin{longtable*}{@{}l @{\quad=\quad} l@{}}
$p(\boldsymbol{x})$  & pressure field \\
% Add a symbol for original/accurate pressure field
$\widetilde{p}(\boldsymbol{x})$ & true pressure field\\
$\boldsymbol{x}$ &    spatial coordinates \\
$\boldsymbol{u}$ &    velocity field \\
$\rho$& density \\
$t$ & time \\
$\mu$ & dynamic viscosity \\
% Add
$k$ & realization number \\
$\bar{p}(\boldsymbol{x})$   & mean of the pressure field\\
% Add
$\mathcal{GP}$ & Gaussian Process\\
$\mathcal{N}$ & Gaussian distribution\\ %
$\mathcal{C}$ & covariance matrix of the Gaussian process\\
$l$ & correlation length in the radial basis function kernel\\
% Add length scale from true pressure field
$l_{p}$ & correlation length scale from the true pressure field\\
$\sigma(\boldsymbol{x})$ & standard deviation of Gaussian Process at a spatial location $\boldsymbol{x}$\\
$\sigma_p$ & standard deviation of the pressure field\\
$\sigma_{\epsilon}$ &  standard deviation of assumed noise level in pressure gradient\\
$\sigma_{\nabla p}$ & standard deviation of embedded noise in pressure gradient\\
$\boldsymbol{O}$ &  vector formed by observations of material derivatives \\
$\boldsymbol{X}_*,\boldsymbol{X}$ &  vector formed by observation locations and general spatial locations \\
$\Sigma_{ij}$ & Covariance matrices\\
$R_{\lambda}$ & Reynolds number based on Taylor micro scale\\
$p_{GPR},p_{ODI}$ & pressure field reconstructed by GPR or ODI
% repeated with previous tilde p
% $p_{T}$ & true pressure field\\ 
\end{longtable*}}

% Introduciton 
\section{Introduction}
\subsection{Pressure field estimation from PIV}

% Poisson Equation Solver
% 
% Previous works on GPR
\lettrine{U}{nderstanding} the pressure field is essential in various turbulence and hydrodynamic researches. The generation of acoustic noise \citep{buchta2017near, wu2008methods} and the onset of boundary separation \citep{gramann1990detection,cerretelli2009boundary}, for example, are instances where an accurate estimation of the pressure is of paramount importance. 
However, non-intrusive measurements of the detailed instantaneous pressure field is a leading challenge in experimental studies. 
By applying Particle Image Velocimetry (PIV), the gradient information of pressure can be obtained from the balance of the Navier-Stokes equation, in which the material acceleration is the dominant term, while the viscous term is negligible at regions away from the wall for high Reynolds number flow \citep{liu2006instantaneous},
\begin{equation}
        \nabla p = - \rho\frac{D \boldsymbol{u}}{Dt} + \mu \nabla^2 \boldsymbol{u} \approx - \rho\frac{D \boldsymbol{u}}{Dt}.
\end{equation}
Numerical tools have been established to further reconstruct the instantaneous pressure field based on the measured pressure gradient. 
While the traditional approach solves the pressure-Poisson equation, which often utilizes an inaccurate boundary condition,
the state-of-art tool, named Omni-directional Integration (ODI), integrates the pressure gradients in a collection of directions and averages the resulting field \citep{liu2003measurements, liu2006instantaneous, liu2016instantaneous, liu2018pressure}. This has led to a robust de-noising framework that greatly mitigates the effect of measurement error \citep{liu2020error}. A recent example shows that the time-averaged pressure reconstructed from Reynolds Averaged Navier-Stokes (RANS) equation based on stereo PIV measurements even reaches spatial resolution beyond that of PIV \citep{moreto2022experimentally}.
The goal of the current study is to extend the idea of denoising and establish a probabilistic framework using Gaussian Process Regression to reconstruct pressure from its gradient information with uncertainties.

\subsection{Gaussian process regression}
% \qw{WRITE A BRIEF INTRODUCTION FOR GPR (Or KRIGING INTERPOLATION)}
% “Gaussian process regression is a powerful, non-parametric Bayesian approach towards regression problems that can be utilized in exploration and exploitation scenarios.” ([Schulz et al., 2018, p. 1]
Gaussian process regression has been used to solve regression problem when the underlying function is unknown and hard to evaluate analytically. 
For example, apply Gaussian Process upper confidence bound sampling (GP-UCB) to provide movie recommendation\cite{schulz2018tutorial}; reconstruct missing temporal and spatial sensor data of a dynamic nonlinear response \cite{ma2022probabilistic,mons2019kriging}; analyze motion trajectory of moving targets from sparse observations \cite{kim2011gaussian}.
This non-parametric Bayesian approach has been proven to be very powerful in exploration and exploitation scenarios. Furthermore, GPR can capture a wide variety of relations between inputs and outputs by the kernel-encoded prior assumption. In this study, we reconstruct the pressure from pressure gradient by deliberately encoding the prior assumption with different kernel functions while previous works mainly focus on obtaining function from observations of the function itself.

The rest of the paper is structured as follows: in \S\ref{sec:formulation} we introduce the mathematical formulation for Gaussian Process Regression when applied to pressure field reconstruction. In \S\ref{sec:problem} we explain how we choose the optimal hyper-parameters for GPR and how we evaluate the performance of GPR and ODI through the forced homogeneous isotropic turbulence database from Johns Hopkins Turbulence Database (JHTDB). In \S\ref{sec:results} we show some comparison results and analyses of reconstructed pressure field by GPR and ODI. In \S\ref{sec:conclusion}, we conclude current results and propose some prospective future works.

\section{Mathematical formulation}
\label{sec:formulation}
In the current study, we propose a new approach, adopting the idea of Gaussian Process Regression (GPR) \cite[see][for a brief tutorial]{schulz2018tutorial}
This probability framework takes into account measurement noise and could help to perform field inversion from gradient information and mitigate the effect of measurement noise.

In Gaussian process regression, we assume the observation of pressure gradient at any spatial location $\boldsymbol{x}$ can be expressed by the true pressure gradient with an additional noise:
\begin{equation}
    \nabla p(\boldsymbol{x}) = \nabla \widetilde{p} (\boldsymbol{x}) + \epsilon
\end{equation}
where the noise term $\epsilon$ follows normal distribution
\begin{equation}
    \epsilon \sim \mathcal{N}(0, \sigma_\epsilon^2).
\end{equation}

Furthermore, GPR regards the pressure field, $p(\boldsymbol{x})$ as a Gaussian process in infinite dimensional space, e.g.
\begin{equation}
    p(\boldsymbol{x}) \sim \mathcal{GP} (\bar{p}(\boldsymbol{x}),\; \mathcal{C}(\boldsymbol{x}, \boldsymbol{x}^{\prime})),
\end{equation}
in which $\bar{p}(\boldsymbol{x})$ represents the mean, or the expected value of the prior distribution for the pressure, $\bar{p}(\boldsymbol{x}) = \mathbb{E}[p(\boldsymbol{x})]$, and $\boldsymbol{x}, \boldsymbol{x'}$ represent two different spatial locations.
The kernel of the Gaussian process, $\mathcal{C}(\boldsymbol{x}, \boldsymbol{x}^{\prime})$ represents the covariance of the uncertain pressure fields at two different spatial locations, $p(\boldsymbol{x})$ and $p(\boldsymbol{x}^{\prime})$. Or equivalently,
$\mathcal{C}(\boldsymbol{x}, \boldsymbol{x}^{\prime}) = \mathbb{E}\left[(p(\boldsymbol{x}) - \bar{p}(\boldsymbol{x}))(p(\boldsymbol{x}^{\prime}) - \bar{p}(\boldsymbol{x}^{\prime}))\right]$.

For a continuous and stationary random process, radial basis function kernel is popularly introduced. The Gaussian kernel, for example, reads
\begin{equation}
\mathcal{C}(\boldsymbol{x}, \boldsymbol{x}^{\prime}) = \sigma(\boldsymbol{x}) \sigma(\boldsymbol{x}^{\prime}) \; \exp({ - \frac 12  \frac{||\boldsymbol{x} - \boldsymbol{x}^{\prime}||^2}{l^2}}),
\end{equation}
in which $\sigma(\boldsymbol{x})$ is the standard deviation and $l$ is the correlation length. 
The standard deviation $\sigma(\boldsymbol{x})$ will be updated during the inference and can naturally lead to uncertainty quantification (UQ) of the reconstruction.  
The correlation length $l$ is a hyper-parameter representing the smoothness of the pressure field, which can be tuned to achieve the best performance for different scenarios.

% Optimize length scale
The corresponding correlation function of the pressure field for the kernel function $\mathcal{C}(\boldsymbol{x}, \boldsymbol{x}^{\prime})$ is defined as
\begin{equation}
    \mathcal{K}(\boldsymbol{x}, \boldsymbol{x}^{\prime}) = \frac{\mathcal{C}(\boldsymbol{x}, \boldsymbol{x}^{\prime})}{\sigma(\boldsymbol{x}) \sigma(\boldsymbol{x}^{\prime})}
\end{equation}
where $\sigma_p$ is the root mean square of entire pressure field.
For radial basis function kernel, the correlation function $\mathcal{K}(\boldsymbol{x}, \boldsymbol{x}^{\prime})$ is only function of distance $r$ between two point,
\begin{equation}
    \mathcal{K}(\boldsymbol{x}, \boldsymbol{x}^{\prime}) = \mathcal{K}(r)
\end{equation}
in which $r = ||\boldsymbol{x} - \boldsymbol{x}^{\prime}||$.
In order to better comprehend the physical meaning of pressure field, we introduce the correlation length scale $l_p$, which could be obtained by fitting the correlation function $\mathcal{K}(r)$ of true pressure field into the correlation function of Gaussian kernel.

\begin{figure}
    \centering
    \includegraphics[width=1\textwidth]{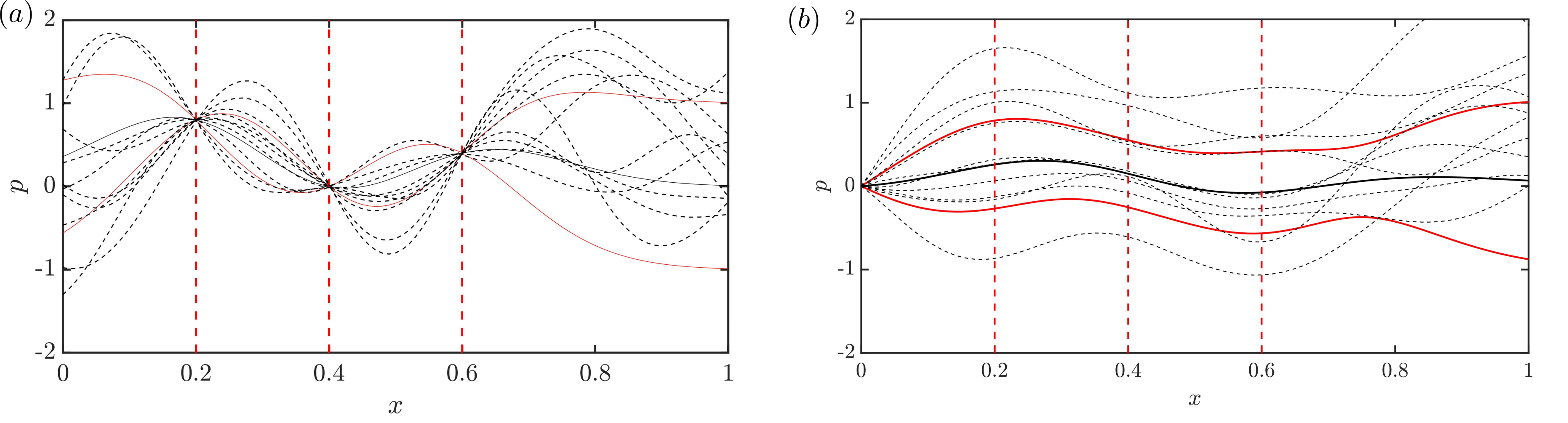}
    \caption{(a): An example of one-dimensional GPR when observations of the values of a smooth function are available. Red dashed lines mark the location of observations while the black dashed lines are samples drawn from the posterior distribution. Two red solid lines mark the region within one standard deviation in the posterior distribution.
    (b): When observations of function derivatives are available, a similar approach can also be adopted to infer the values of the function.}
    \label{fig:1D_examples}
\end{figure}
    
While most applications of GPR deal with observing the values of the function directly \citep{mons2019kriging}, as shown in Figure \ref{fig:1D_examples}$(a)$, the formulation of GPR is general and can be applied to observations of the gradient information. An example of such reconstruction for a one-dimensional case is shown in Figure \ref{fig:1D_examples}$(b)$, where the dashed lines marks the observation location for the gradient of the function $p(x)$. 
    Given $\boldsymbol{X}_*$ a vector collection of $n$ spatial locations $\boldsymbol{x_n}$ where data are observed, the samples of the pressure gradient observations would follow a multivariate Gaussian distribution,
    \begin{equation}
        \boldsymbol{O} = \nabla p\left(\boldsymbol{X}_*\right) \sim \mathcal{N}\left( \nabla \bar{p}\big|_{\boldsymbol{X}_*}, \nabla_{\boldsymbol{x}}\nabla_{\boldsymbol{x}^{\prime}} \mathcal{C} \left(\boldsymbol{X}_*,\boldsymbol{X}_*\right) + \sigma_{\epsilon}^2 \mathbf{I}_* \right).
    \end{equation}
    Here $\sigma_{\epsilon}$ is the assumed noise level of synthetic noise introduced as a mimic of experimental data. $\mathbf{I}_*$ is the identity matrix. 
    Moreover, once the observations are drawn from the above distribution, the observations of pressure gradient at $\boldsymbol{X}_*$ and unknown values of pressure field at $\boldsymbol{X}$, a vector collection of spatial locations where reconstruction are conducted, would follow the joint Gaussian distribution,
    \begin{equation}
        \begin{bmatrix}
            \boldsymbol{O}\\\\
            p(\boldsymbol{X})
        \end{bmatrix} \sim \mathcal{N}\left(
        \begin{bmatrix}
            \nabla p\left(\boldsymbol{X}_*\right)\\\\\\
            \bar{p}(\boldsymbol{X})
        \end{bmatrix}, 
        \begin{bmatrix}
        \underbrace{\nabla_{\boldsymbol{x}}\nabla_{\boldsymbol{x}^{\prime}} \mathcal{C} \left(\boldsymbol{X}_*,\boldsymbol{X}_*\right) + \sigma_{\epsilon}^2 \mathbf{I}_*}_{\Sigma_{11}},  \underbrace{\nabla_{\boldsymbol{x}^{\prime}} \mathcal{C} \left(\boldsymbol{X},\boldsymbol{X}_*\right)}_{\Sigma_{12}}\\
            \underbrace{\nabla_{\boldsymbol{x}} \mathcal{C} \left(\boldsymbol{X}_*,\boldsymbol{X}\right)}_{\Sigma_{21}}, \underbrace{\mathcal{C} \left(\boldsymbol{X},\boldsymbol{X}\right)}_{\Sigma_{22}}
        \end{bmatrix}
        \right)
    \end{equation}
    The last piece of this joint Gaussian distribution is the reference pressure.
    Since the value of reference pressure does not influence the dynamic structure of reconstructed pressure fields, we include one additional observation of pressure $p(\boldsymbol{x}_0)=0$ at location $\boldsymbol{x}_0$ in the formulation.
    Once noisy measurements for the pressure gradient become available, the pressure field can be recovered using Bayes' theorem. And the posterior conditional distribution is given by,
    \begin{equation}
        p\left(\boldsymbol{X}\right) \sim \mathcal{N}\left( \bar{p} -\Sigma_{21}\Sigma_{11}^{-1} \left(\boldsymbol{O}-\nabla\bar{p}\big|_{\boldsymbol{X}_*}\right),\Sigma_{22} - \Sigma_{21} \Sigma_{11}^{-1}\Sigma_{12}\right).
    \end{equation}
    The updated mean of the posterior distribution is regarded as the reconstructed pressure field from GPR,
    \begin{equation}
        p_{GPR}(\boldsymbol{X}) = \bar{p} -\Sigma_{21}\Sigma_{11}^{-1} \left(\boldsymbol{O}-\nabla\bar{p}\big|_{\boldsymbol{X}_*}\right)
    \end{equation}
    
    Notice that the matrix inversion in the above expression requires $\mathcal{O}(N^3)$ operations, with $N$ being the number of observations.
    The inversion would therefore be computationally intractable for large $N$. 
    Nevertheless, for large-scale computations, we could transform the matrix inversion into an iterative algorithm using the Conjugate Gradient method \citep{hestenes1952methods}, which is part of future efforts.

Moreover, the covariance matrix of the posterior probability distribution can be computed from,
\begin{equation}
C_{GPR}(\boldsymbol{X}, \boldsymbol{X}) = \Sigma_{22} - \Sigma_{21} \Sigma_{11}^{-1}\Sigma_{12}.
\end{equation}
Square roots of the diagonal elements of the above matrix are the standard deviation $\sigma_p(\boldsymbol{X})$, representing the uncertainty of the reconstructed pressure fields at different spatial locations.

\section{Problem Setup}
\label{sec:problem}
\begin{figure}
\centering
\begin{minipage}{.45\linewidth}
    \includegraphics[width=\textwidth]{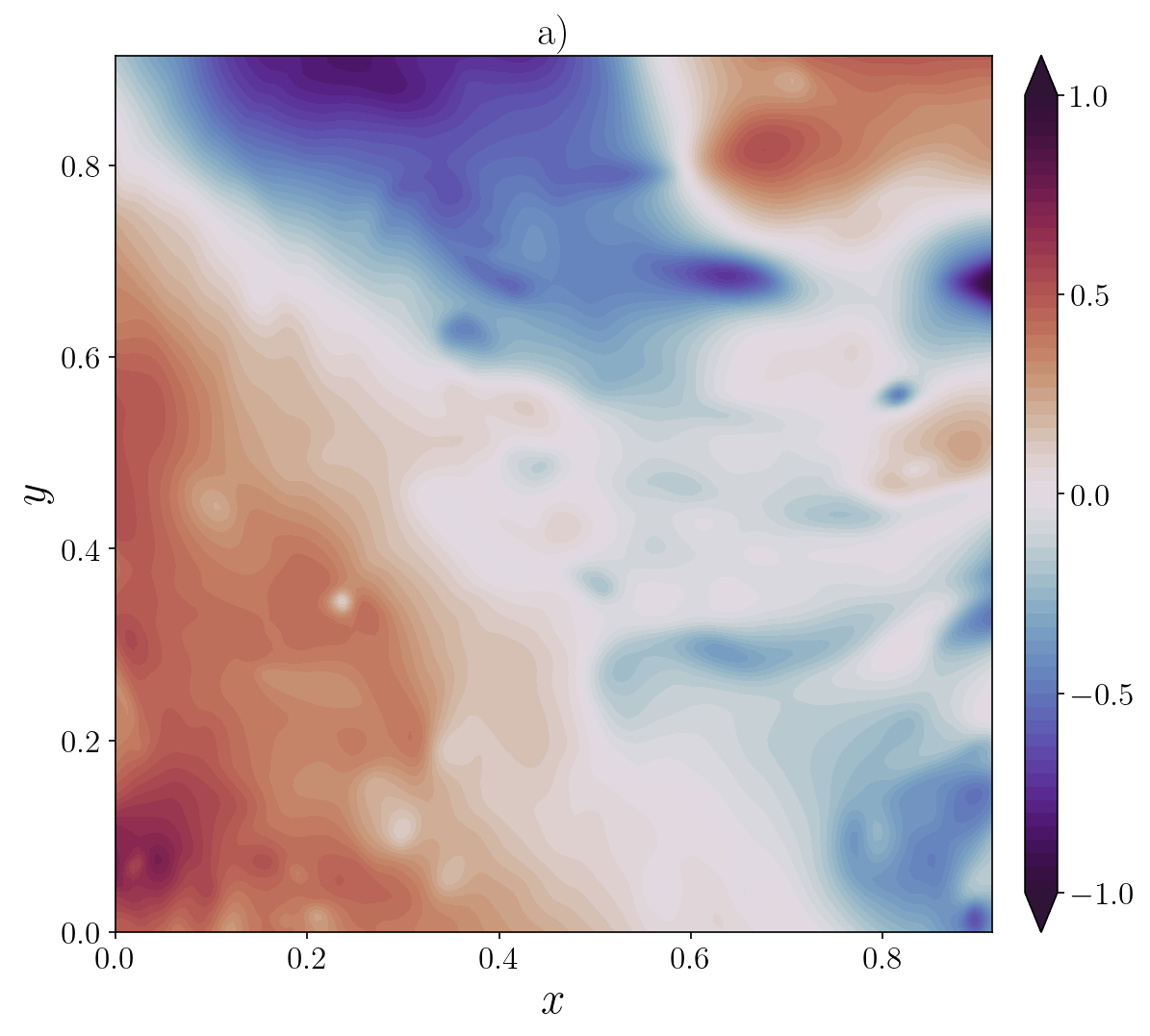}
\end{minipage}\quad%
\begin{minipage}{.45\linewidth}
    \includegraphics[width=\textwidth]{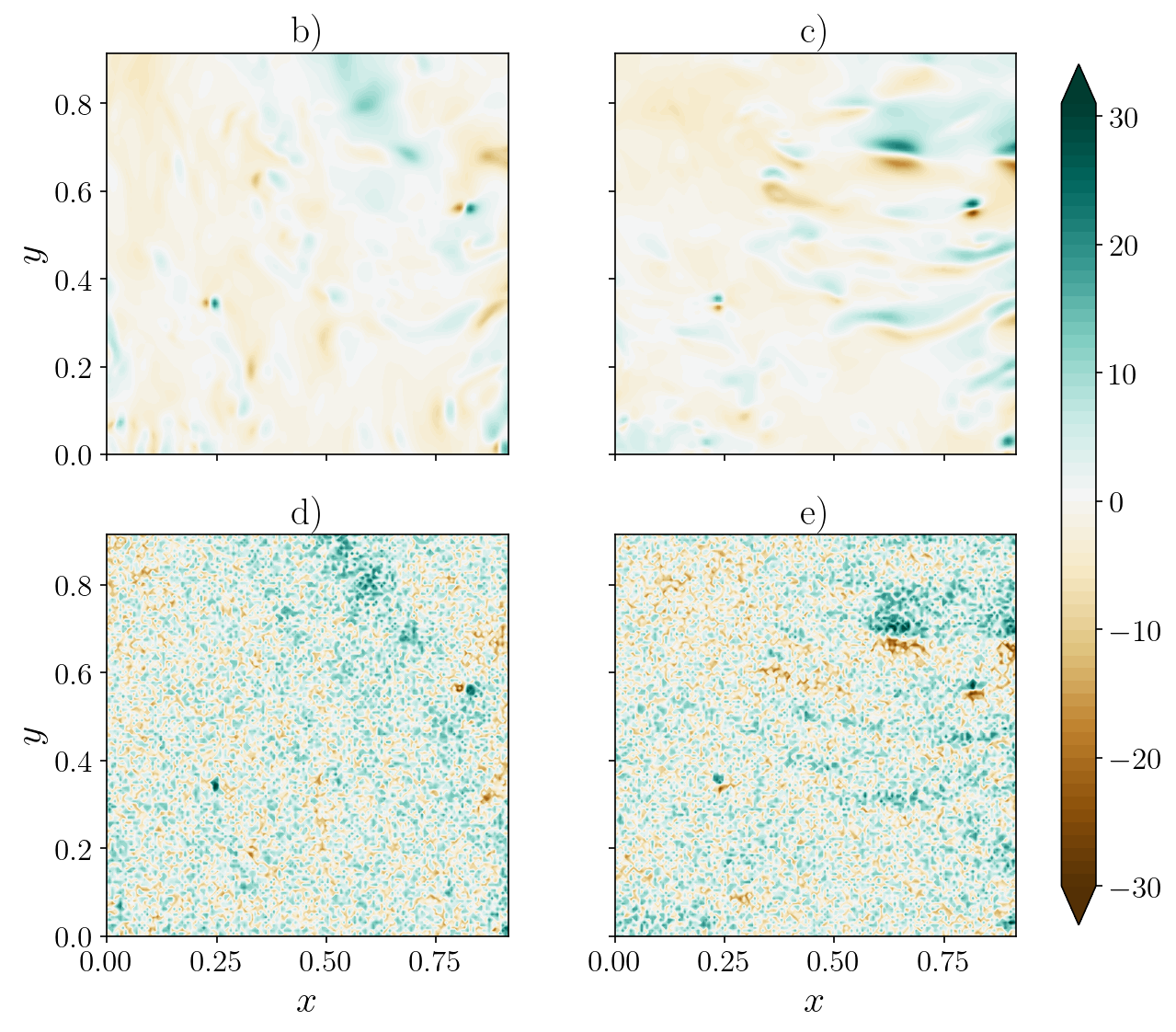}
\end{minipage}%
\caption{(a): True pressure field from an isotropic turbulence DNS database. (b)-(c): Pressure gradient obtained from the DNS pressure field by central finite difference method. (d)-(e): Sample realization of 1000 error embedded pressure gradient.}
\label{fig:problem}
\end{figure}
\subsection{Data acquisition}
As the first step, we test our algorithm in a homogeneous isotropic turbulence flow field with Reynolds number around $R_\lambda \sim 433$ based on Taylor microscale. 
Instead of using data from experiments, we extract data in the Johns Hopkins Turbulent database \citep{li2008public,perlman2007data,yeung2012dissipation} from a direct numerical simulation (DNS) as a surrogate. The simulated data provides access to the fully resolved velocity and pressure fields, enabling us to test our algorithm thoroughly.
Although the database is three-dimensional, the algorithm of both ODI and GPR is able to reconstruct pressure on a two-dimensional plane given observations of the in-plane components of the gradient vectors, as shown in Figure \ref{fig:problem}.

The observations are the projection of material derivatives onto the $x-y$ plane at certain observation locations extracted from the database, at four times the DNS grid spacing in $x$ and $y$ directions.
A total number of $150 \times 150$ observations of pressure gradient are obtained on a two-dimensional plane in the turbulent field on the $x-y$ plane with $L_x = L_y = 0.9$.

In order to compare the performance of GPR and ODI, we adopted 1000 realizations of error-embedded pressure gradient generated by \citet{liu2020error} and reconstruct the pressure field by ODI as well as GPR. 
% added error description
The error-embedded pressure gradient is generated by adding random noise of uniform distribution with the magnitude as 40\% of $(|\nabla p|_{\text{DNS}})_{\text{max}}$, the maximum magnitude of the true pressure gradient, at each point.
The true pressure field, true pressure gradient as well as one sample realization of error embedded pressure gradient are shown in Figure \ref{fig:problem}.

\subsection{Evaluation of pressure reconstruction}
To quantify the accuracy of pressure reconstruction when subject to noise in the measurements, as often observed in PIV, we evaluate the cumulative error over 1000 realizations of noisy measurements. 
First, we calculate the error of the reconstructed pressure field by GPR and ODI by subtracting the true pressure field $\widetilde{p}(\boldsymbol{x})$ at each point, i.e.,
\begin{equation}
    \epsilon_{ij} = p_{ij} - \widetilde{p}_{ij},
\end{equation}
where $i$, $j$ refer to Cartesian indices. Then, the standard deviation of error $\epsilon_{\text{std}}$ is defined as
\begin{equation}
    \epsilon_{\text{std}} = \sqrt{\frac{1}{N_x\times N_y -1} \sum_{j=1}^{N_y}\sum_{i=1}^{N_x} (\epsilon_{ij} - \overline{\epsilon_{ij}})^2}.
\end{equation}
Finally, we calculate the averaged standard deviation over 1000 realizations, the cumulative error $\varepsilon_{\text{std}}$, as 
\begin{equation}
    \varepsilon_{\text{std}} = \frac{1}{k}\sum_{n=1}^{k} \left(\frac{\epsilon_{\text{std}}}{p_{\text{std}}}\right)_n.
\end{equation}

% The correlation length is set to be $l = 0.0625$ and the noise level $\sigma_{\epsilon} = 6$.

% Preliminary results demonstrate good agreement between the ODI and GPR with the true pressure field.
% As a further quantification, we ploted the distribution of errors in the reconstruction for ODI and GPR in Figure \ref{fig:compare_GPR_ODI}. 
% A better reconstruction can be achieved using GPR, with nearly half of the error compared with ODI.

% To further investigate the de-noising performance of the GPR algorithm, similar to Liu and Moreto (2020)\citep{liu2020error}, the GPR algorithm will be further tested with 1000 sample realizations of pressure gradient distribution of the isotropic turbulence field embedded with statistically independent random noise.  The random noise has standard uniform spatial distributions and zero mean value, and is generated using a built-in Matlab® function ‘rand’, with 1000 distinct seed numbers. The noise amplitude
% is set to 40$\%$ of the maximum magnitude
% of the pressure gradient in the sample DNS isotropic turbulence field. Statistical results based on these 1000 sample realizations will be compared with those obtained in Liu and Moreto (2020)\citep{liu2020error} and presented in the final paper.

\subsection{Hyper-parameter Optimization}
% Normalized
\begin{figure}[!h]
    \centering
    \includegraphics[width=\textwidth]{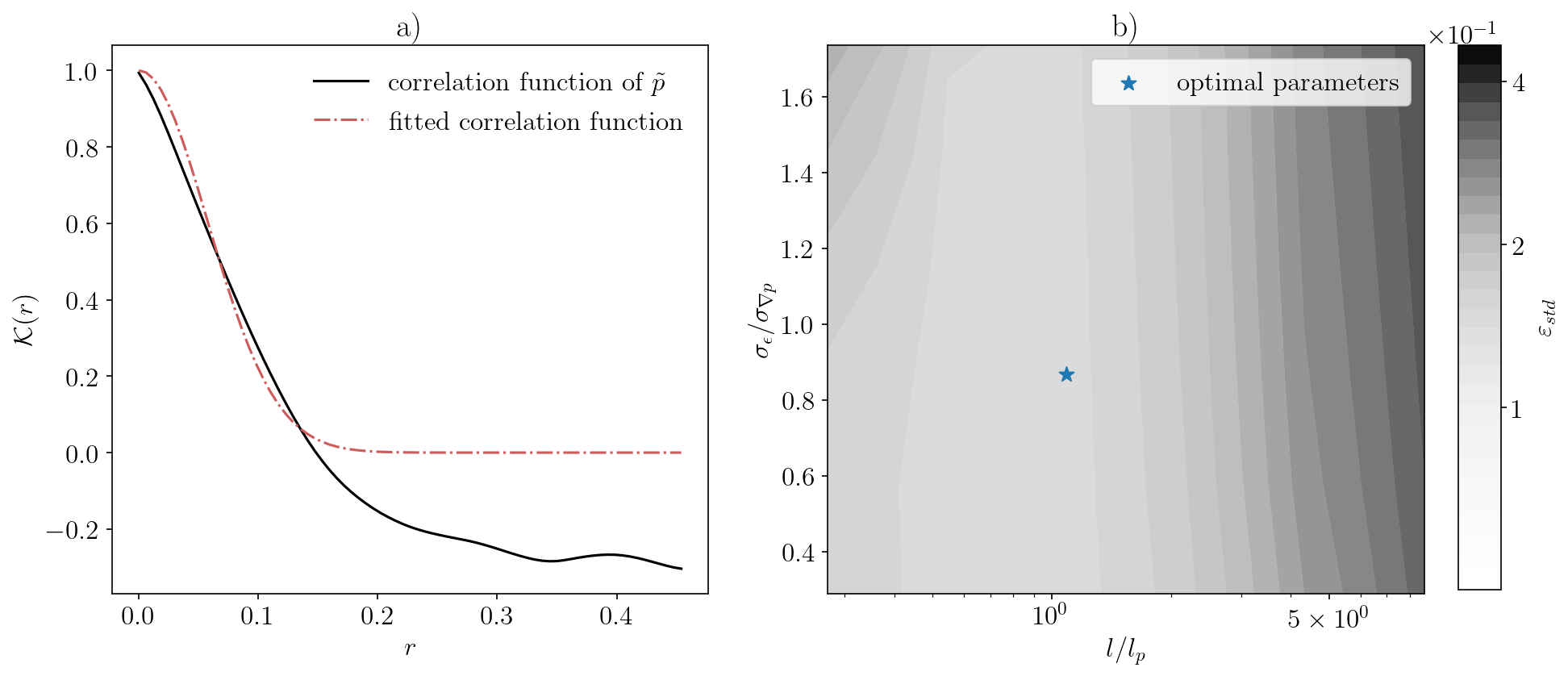}
    \caption{(a): Correlation function $K(r)$ of the true pressure field and the optimal Gaussian kernel function from curve-fitting. (b): The cumulative error of 10 realizations with different correlation length $l$ and assumed noise level $\sigma_{\epsilon}$ as well as the location of optimal hyper-parameters.}
    \label{fig:hyper_optimize}
\end{figure}
The performance of both ODI and GPR methods depends on setting up proper parameters, or hyper-parameters.
For Omni-directional Integration, we adopt the state-of-art formulation with Rotating Parallel Rays to create different integration paths \citep[see][for details]{liu2016instantaneous}.
The parallel rays are rotated with an increment of 0.2 degrees, which creates a total number of 1800 different orientations of parallel integration paths over the entire domain of calculation. The optimal separation between adjacent parallel rays at 0.4 times of the grid size as suggested by \citep{liu2020error} is adopted in the ODI calculation. 
For GPR, the prior distribution is set with mean $\bar{p}(\boldsymbol{x}) = 0$ and variance $\sigma(\boldsymbol{x}) = 1$. However, correlation length $l$ and assumed noise level $\sigma_\epsilon$ still need to be optimized, which is going to be elaborated in the following sections. 

In order to optimize the performance of GPR, we calculate the averaged error of 10 realizations with different hyper-parameters sets and find the optimal hyper-parameters with the lowest error as the correlation length $l=0.0625$ and the assumed noise level $\sigma_\epsilon=6$. 
The error is slightly smaller with assumed noise level $\sigma_\epsilon$ less than 6. However, because the influence of equivalent noise level is not very significant once it is smaller than 6, we choose $\sigma_\epsilon =6$ as our optimal hyper-parameter.
The correlation length scale $l_p$ could be obtained by fitting the Gaussian kernel into the correlation function of the true pressure field, which is equal to 0.0574 for the isotropic turbulence. So the correlation length scale $l_p = 0.0574$. Embedded noise is a uniform distribution from -12 to 12, the standard deviation of which is 6.928. Thus the magnitude of embedded noise is $\sigma_{\nabla p}=6.918$.
We could find out that optimal hyper-parameters correlation length $l$ approximates to correlation length scale $l_p$ while assumed noise level $\sigma_\epsilon$ approximates to the magnitude of embedded noise $\sigma_{\nabla p}$. This phenomenon proves that the kernel of GPR depends on the correlation function of the true pressure field as well as the standard deviation of noise in the observations.

\section{Results}
\label{sec:results}
\subsection{Error Analysis in Physical Space}
\begin{figure}
    \centering
    \includegraphics[width=0.7\textwidth]{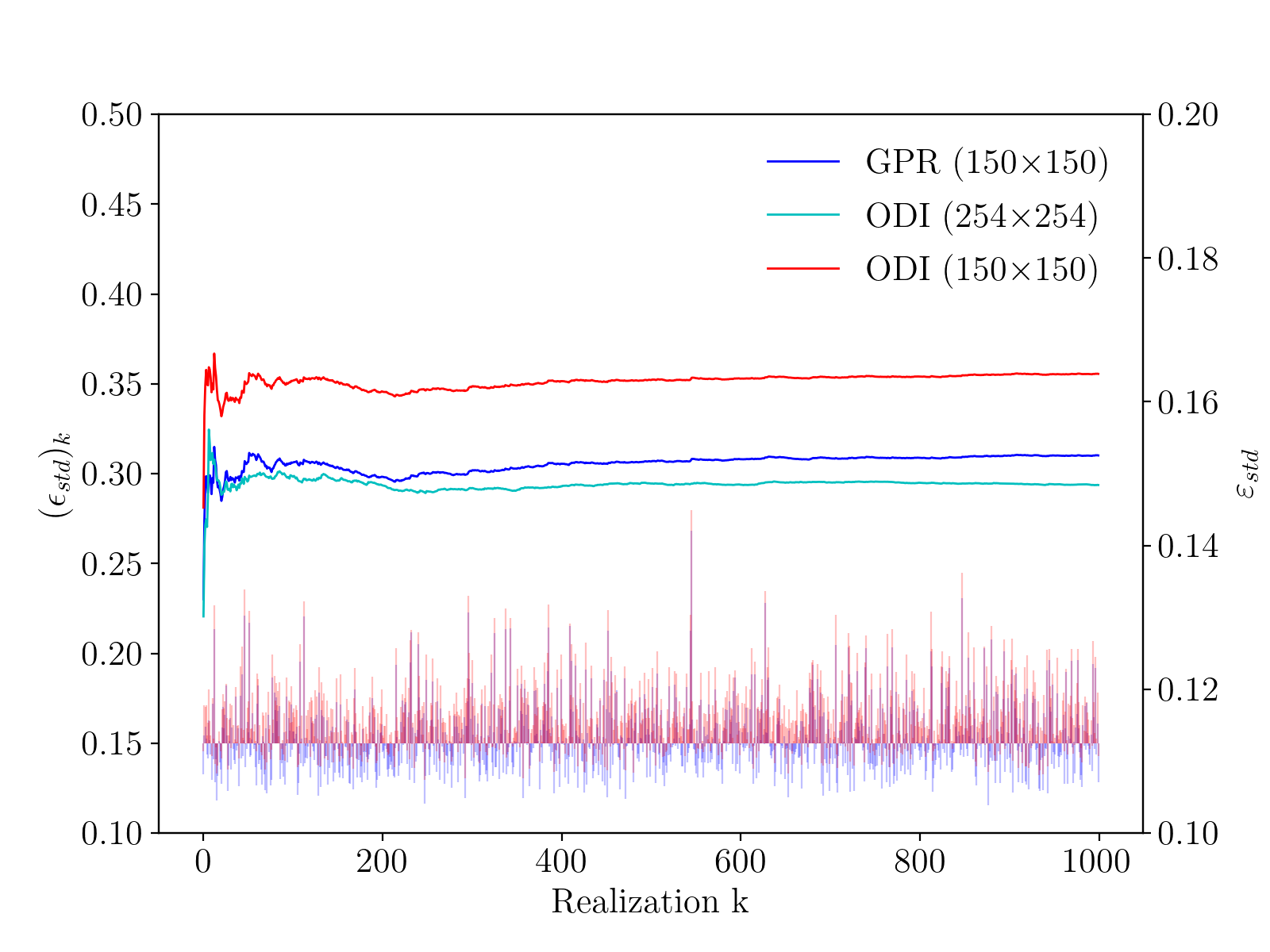}
    \caption{Cumulative error $\varepsilon_{\text{std}}$ and standard deviation $\left(\epsilon_{\text{std}}\right)_k$ of 1000 realizations with correlation length $l=0.0625$ and assumed noise level $\sigma_\epsilon=6$. The bottom histogram shows the standard deviation of error $(\epsilon_{\text{std}})_k$ in the $k$-th realization. The blue bar represents the standard deviation of error by GPR and red line represents the standard deviation of error by ODI in $150\times150$ grids. Three converged lines on the top show the cumulative error of GPR and ODI in $150\times150$ and $254\times254$ grids.}
    \label{fig:realizations}
\end{figure}

% Reconstructed Accuracy comparison between ODI and GPR
The cumulative error of GPR and ODI on 150 by 150 grids as well as the cumulative error of ODI on 254 by 254 grids are shown in Figure \ref{fig:realizations}. From the result, we could observe that error of GPR converges to 0.153 over 1000 realizations. Furthermore, GPR has a similar level of accuracy with ODI method on 150 by 150 grids domain. However, ODI on 254 by 254 grids can reach to lower cumulative error 0.148, the same as the result of previous work by \citet{liu2020error}. But GPR requires more memory to conduct the matrix inversion in the regression.
% , so we are not able to conduct the comparison on the larger domain for now.
Therefore, for large-scale problems, it is necessary to convert matrix inversion into an iterative algorithm, such as the conjugate gradient iteration for better efficiency, which is part of future work.

% Worst Scenerio
\begin{figure}
    \centering
    \includegraphics[width=\textwidth]{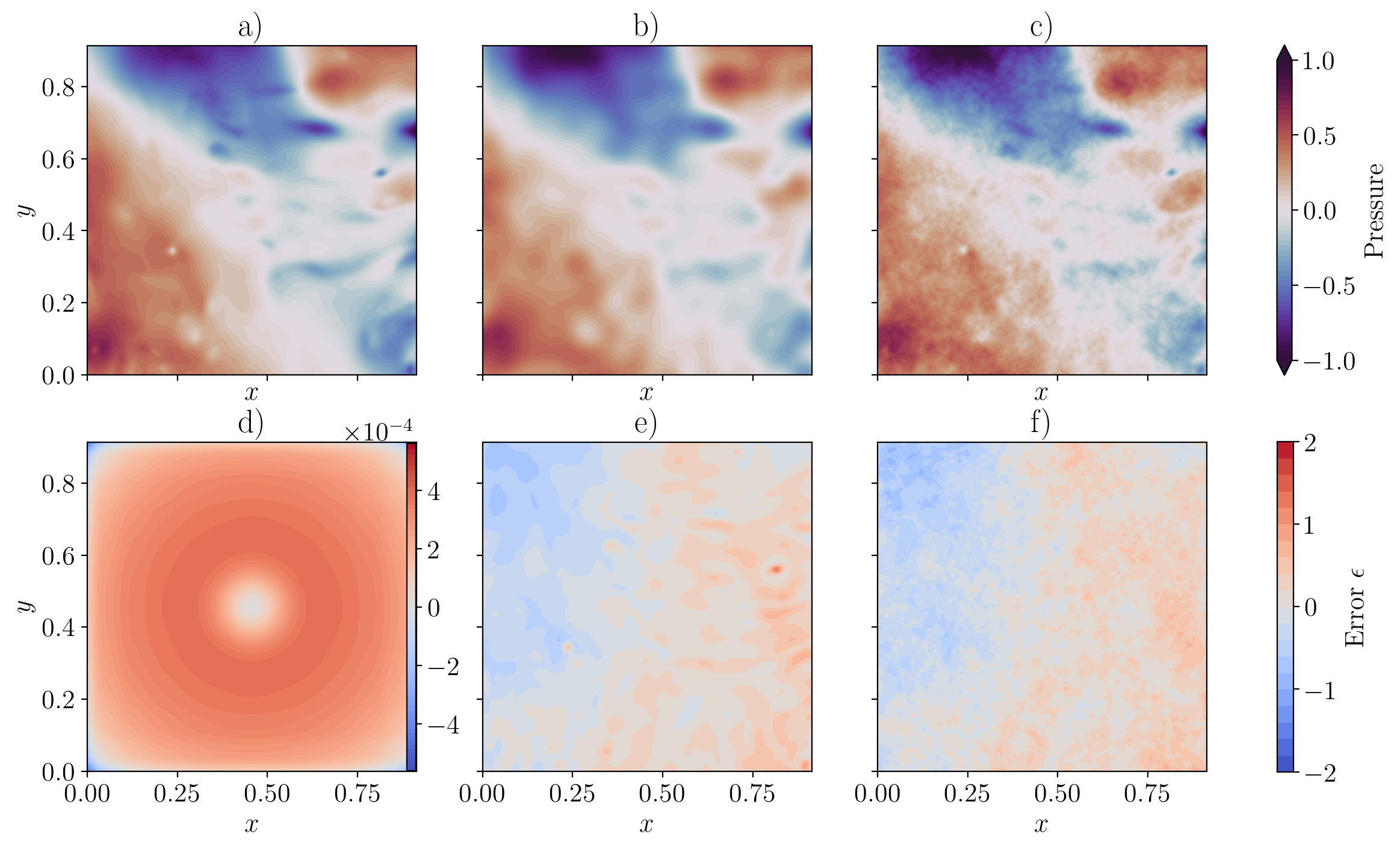}
    \caption{(a): True pressure field from isotropic turbulence DNS database. (b): Instant of realization of reconstructed pressure field by GPR with the largest standard deviation of error $\epsilon_{\text{std}}$. (c): Instant of realization of reconstructed pressure field by ODI with the largest standard deviation of error $\epsilon_{\text{std}}$. (d): Standard deviation of reconstructed pressure field by GPR, $\sigma(\boldsymbol{X})$ subtracted by the standard deviation of reference point $\sigma(\boldsymbol{x}_0)$. (e): Error distribution of reconstructed pressure field by GPR,  obtained by subtracting (a) from (b). (f): Error distribution of reconstructed pressure field by ODI,  obtained by subtracting (a) from (c).}
    \label{fig:comparison}
\end{figure}

To further investigate the performance of GPR and ODI, Figure \ref{fig:comparison} shows the exact pressure field $\widetilde{p}$, the pressure field reconstructed by GPR for the instance of the  worst case scenario among the 1000 realizations, pressure field reconstructed by ODI for the instance of the  worst case scenario among the 1000 realizations, the standard deviation of reconstructed pressure field by GPR, error of GPR result as well as error of ODI result of the worst performance case among 1000 realizations. In Figure \ref{fig:comparison}, we could observe that the reconstructed pressure field by GPR is significantly smoother than ODI result and both reconstructed pressure fields are fairly accurate compared to the exact pressure field. Although the reconstructed pressure field by GPR has a smaller global error, it also has a larger local error compared to the reconstructed pressure field by ODI, indicating the tendency of GPR in flattening impulsive local pressure changes. Furthermore, the reconstructed pressure field by ODI preserves more fine structures (combined with noise and high-frequency pressure signal) while GPR seems to have a stronger denoising effect according to the smooth distribution. 

%\subsection{\del{Variance}}
% Varaince of reconstructed posterior distribution
While the mean of the posterior distribution can be used to represent the reconstructed pressure field, we could also evaluate the accuracy of reconstruction based on the standard deviation of the posterior distribution. 
Notice that adding or subtracting a constant field on the pressure does not alter the agreement with the observation of the pressure gradient.
This property, sometimes coined as ``gauge invariance", has to be eliminated in order to obtain meaningful results for uncertainty analysis.
Therefore, we assume that the reconstructed pressure at a given reference point, e.g. $\boldsymbol{x}_0$ at the center of the domain is solely due to such freedom of adding an arbitrary constant field. 
The variances of the pressure field at other locations further take into account the effect of $\sigma_p(\boldsymbol{x}_0)$, which should be subtracted to obtain a reasonable estimation of the reconstruction uncertainty without the influence of such gauge invariance.
For this reason, we here plot $\sigma_p(\boldsymbol{X}) -\sigma_p(\boldsymbol{x}_0)$ in Figure \ref{fig:comparison}(d).

% \begin{figure}
% \centering
% \begin{minipage}{.45\linewidth}
%     \includegraphics[width=\textwidth]{Figures/sample-pressure.png}
% \end{minipage}\quad%
% \begin{minipage}{.45\linewidth}
%     \includegraphics[width=\textwidth]{Figures/variance.png}
% \end{minipage}%
% \caption{(a): Mean of the posterior distribution, which is reconstructed pressure field by GPR. (b): Variance of the posterior distribution, which is the variance of the reconstructed result.}
% \label{fig:variance}
% \end{figure}

\begin{figure}
    \centering
    \includegraphics[width=0.8\textwidth]{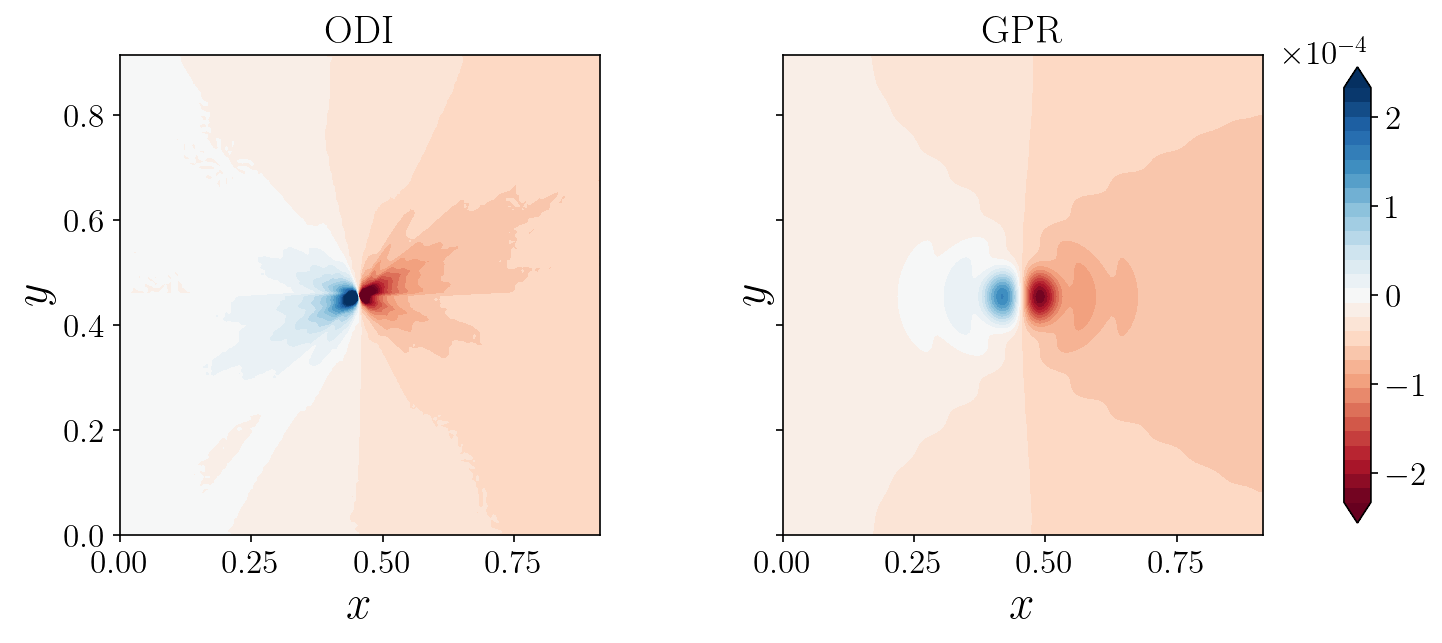}
    \caption{Error Propagation of GPR and ODI, represented by the change of reconstructed pressure field when perturbation in the form of a unit impulse of $\partial p \partial x$ is added at the center of the computational domain.}
    \label{fig:error-propagation}
\end{figure}

% Error Propagation
%\subsection{Impulse-response analysis}
In order to visualize how the error is propagated throughout the entire pressure field by different methods, we add a unit impulse in $\frac{\partial p}{\partial x}$ at the middle of the computational domain and use GPR and ODI to reconstruct the pressure field, respectively. 
Differences between the reconstructed pressure field with and without the unit impulse are then visualized in Figure \ref{fig:error-propagation} to quantify the domain of influence for such point-wise perturbation.
Such approaches have been adopted in previous research to study the domain of influence or the domain of dependence to fully understand the forward and backward propagation of perturbations in a dynamical system or data assimilation algorithm \citep{wang_wang_zaki_2022,wang2019spatial}.

The results of this impulse response have profound implications.
First of all, the ODI method indicates a clear singularity point at the location of perturbation, manifested by very large positive and negative values.
The implication here is that although ODI averages the error across the whole computational domain, the reconstructed pressure still relies heavily on the local pressure gradient information near the point of interest. In other words, most of the local error remains local, and correspondingly, the error diffusion is relatively not strong in comparison with that of GPR.
On the other hand, the GPR method exhibits nearly zero influence at the point of perturbation and has a larger influence slightly farther away.
This difference could explain the stronger de-noising effect in GPR than in ODI.

\subsection{Error Analysis in Wave Number Space}
% Energy Spectrum
\begin{figure}
    \centering
    % \begin{minipage}{.45\linewidth}
    % \includegraphics[width=\textwidth]{Figures/energy-spectrum-density.png}
    % \end{minipage}\quad%
    % \begin{minipage}{.45\linewidth}
    %     \includegraphics[width=\textwidth]{Figures/energy-spectrum-density-withournoise.png}
    % \end{minipage}%
    \includegraphics[width=\textwidth]{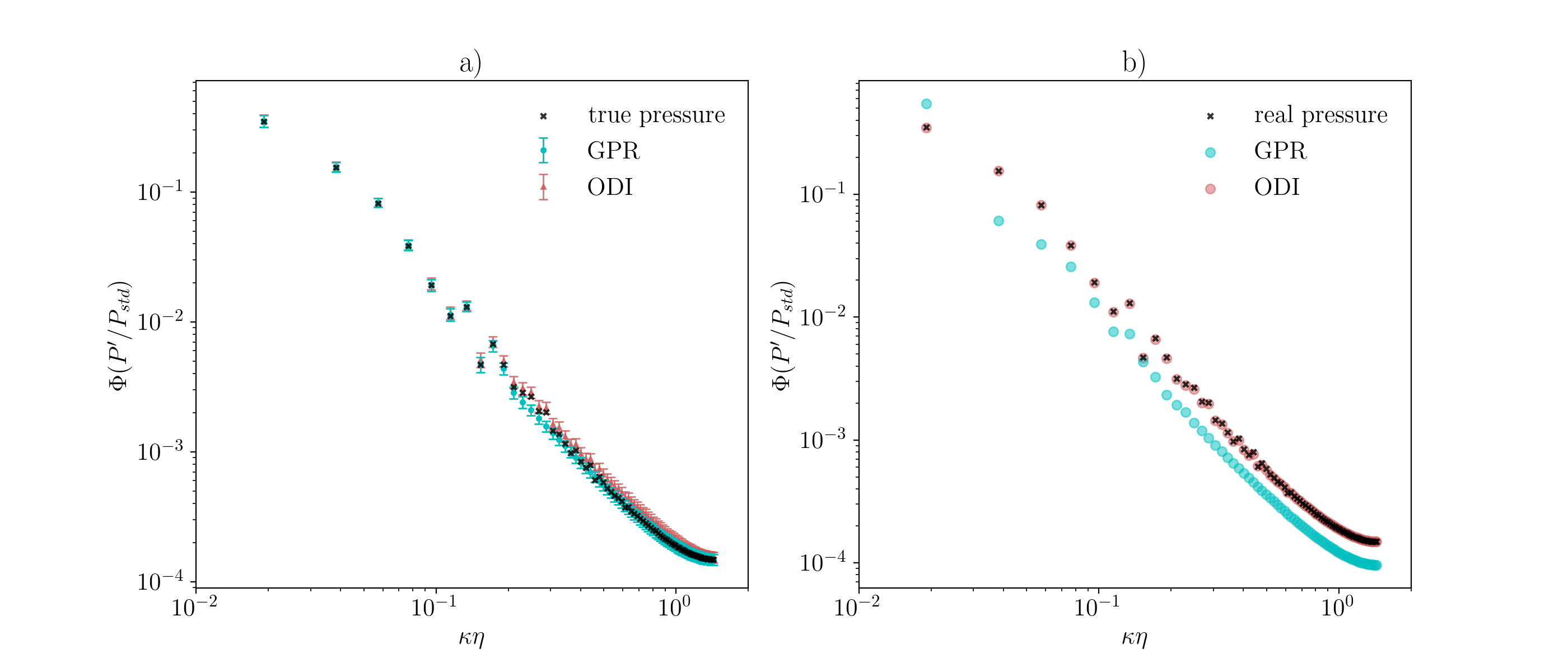}
    \caption{(a): Power spectrum density of true pressure field as well as reconstructed pressure field by GPR and ODI from error embedded pressure gradients of 150 by 150 grids. (b): Power spectrum density of true pressure field as well as reconstructed pressure field by GPR and ODI from true pressure gradients of 150 by 150 grids}
    \label{fig:power-spectrum-density}
\end{figure}
The different behavior of GPR and ODI in Figure \ref{fig:comparison} leads to a comparison of power spectrum density of the reconstructed pressure field by GPR and ODI, as shown in Figure \ref{fig:power-spectrum-density}. 
From Figure \ref{fig:power-spectrum-density}(a), we could see that both methods perform well on low wave number space: the power spectrum density of GPR and ODI coincides with the power spectrum density of true pressure field perfectly. 
However, in high wave number space, the power spectrum density of different pressure field seem to diverge: GPR has a lower power spectrum density while ODI has a larger power spectrum density compared to the power spectrum density of true pressure field. 
This phenomenon validates previous observations in the pressure field shown in Figure \ref{fig:comparison}, as well as the impulse response in Figure \ref{fig:error-propagation}.
GPR has a stronger denoising effect but also smooths out some pressure information in high wave number space while ODI preserves more fine structures. 
Some of them are dynamic behavior of the pressure field, others are noise in the observation.
Furthermore, Figure \ref{fig:power-spectrum-density}(b) clearly indicates that if the pressure gradient information is accurate, ODI can faithfully replicate the dynamic behavior of the pressure field over the entire spectral range. However, in contrast, because the hyperparameters used in this GPR computation were optimized with the error-embedded data (therefore are flow dependent), GPR produces a reconstructed pressure spectrum with an overall lower fluctuation amplitude over the entire spectral domain. This indicates that the optimization of GPR needs to be adjusted  according to the actual flow properties.  
This problem might be solved by switching a proper kernel function rather than a Gaussian kernel in this case since its denoising effect is too strong to preserve necessary information, which requires future work.

% \begin{figure}
% \centering
% \begin{minipage}{.45\linewidth}
%     \includegraphics[width=\textwidth]{Figures/error_propagation_150_GPR.png}
% \end{minipage}\quad%
% \begin{minipage}{.45\linewidth}
%     \includegraphics[width=\textwidth]{Figures/error_propagation_150_ODI.png}
% \end{minipage}%
% \caption{Error Propagation of GPR and ODI}
% \label{fig:error-propagation}
% \end{figure}

% Problem 
% 1. The variance of entire domain is pretty large while on a smaller domain, the variance is almost 0
% 2. The circle in the middle is created by the location of reference pressure. 

% -----------------------------

\section{Conclusion and future work}
\label{sec:conclusion}
We adopt the framework of Gaussian Process Regression (GPR) to the problem of determining the pressure fields from measured pressure gradients, with the potential of reconstructing pressure from sparsely measured data. 
The formulation naturally avoids the burden of solving Poisson equation with inaccurate boundary conditions and takes into account the effect of measurement noise.
Furthermore, this framework provides more possibilities to improvement.

From the comparison between the reconstructed pressure field by GPR and ODI, we are able to conclude that the reconstructed pressure field by GPR achieves accuracy comparable to the state-of-art Omni-directional integration method (ODI) and has a stronger denoising effect compared to ODI. 
However, pressure reconstruction by GPR might also smooth out some pressure information in high wave number space by mistake, especially dealing with accurate pressure gradient data.
This problem might be able to be solved by switching the Gaussian kernel, which requires further investigation.

In future directions, the following will be studied in detail: (a) improvement of computational efficiency for large-scale problems. (b) different kernel functions for the Gaussian Process. (c) effect of the sparseness of the observation in terms of reconstruction quality.

% acknowledgement
\section{Acknowledgments}
The support from San Diego State University is gratefully acknowledged. Thanks to Jose Moreto for his assistance in the usage of the parallel-ray Omni-directional integration code and for providing the error-embedded pressure gradients data in previous study by \citet{liu2020error}.

\bibliography{main}

\end{document}